\documentclass[review,authoryear]{elsarticle}
\usepackage{graphicx,afterpage}
\usepackage{bm}
\usepackage{soul}
\usepackage{empheq}
\usepackage{mathtools}
\usepackage{algorithm,algpseudocode}
\usepackage[margin=1.25in]{geometry}
\usepackage{mathrsfs,color}
\usepackage{amsfonts}\usepackage{multirow}
\usepackage{amssymb}\usepackage{caption,comment}
\usepackage{amsmath}\usepackage{float}
\usepackage{amssymb,amsthm}
\usepackage{graphicx,afterpage,cancel}
\usepackage{epstopdf}
\usepackage{natbib}
\usepackage{caption}
\usepackage{subcaption}
\usepackage{float}
\usepackage{enumitem}
\usepackage{hyperref}
\hypersetup{CJKbookmarks,%
bookmarksnumbered,%
colorlinks,%
linkcolor=black,%
citecolor=black,%
plainpages=false,%
pdfstartview=FitH,
pdfauthor=author}

\definecolor{orange}{RGB}{255,127,0}

\usepackage{soul}
\usepackage{epstopdf}
\graphicspath{ {./Figs/}}
\afterpage{\clearpage}

\begin{document}

\begin{frontmatter}

\title{A Physics-Informed Variational Inference Framework for Identifying Attributions of Extreme Stress Events in Low-Grain Polycrystals}

\author[1]{Yinling Zhang}
\author[2]{Samuel D. Dunham}
\author[2]{Curt A. Bronkhorst}
\author[1]{Nan Chen\corref{cor1}}
\ead{chennan@math.wisc.edu}

\cortext[cor1]{Corresponding author}

\address[1]{Department of Mathematics, University of Wisconsin, Madison, WI 53706, USA}
\address[2]{Department of Mechanical Engineering, University of Wisconsin, Madison, WI 53706, USA}

\begin{abstract}
Polycrystalline metal failure often begins with stress concentration at grain boundaries. Identifying which microstructural features trigger these events is important but challenging because these extreme damage events are rare and the failure mechanisms involve multiple complex processes across scales. Most existing inference methods focus on average behavior rather than rare events, whereas standard sample-based methods are computationally expensive for high-dimensional complex systems.
In this paper, we develop a new variational inference framework that integrates a recently developed computationally efficient physics-informed statistical model with extreme value statistics to significantly facilitate the identification of material failure attributions. First, we reformulate the objective to emphasize observed exceedances by incorporating extreme-value theory into the likelihood, thereby highlighting tail behavior.
Second, we constrain inference via a physics-informed statistical model that characterizes microstructure-stress relationships, which uniquely provides physically consistent predictions for these rare events.
Third, mixture models in a reduced latent space are developed to capture the non-Gaussian characteristics of microstructural features, allowing the identification of multiple underlying mechanisms.
In both controlled and realistic experimental tests for the bicrystal configuration, the framework achieves reliable extreme-event prediction and reveals the microstructural features associated with material failure, providing physical insights for material design with uncertainty quantification.

\end{abstract}

\end{frontmatter}

\section{Introduction}

Metallic material failure often originates from extreme stress events, in which local stresses exceed critical thresholds, triggering the nucleation of voids and catastrophic damage. Plasticity mechanisms in these events create highly nonuniform stress and strain fields at grain and sub-grain scales \cite{bronkhorst2007modeling, lieberman2016microstructural, bronkhorst2021local, schmelzer2025statistical, zhang2023data}. These heterogeneous fields reflect the underlying microstructure, meaning certain grain- or boundary-level features can trigger extreme stress localization. Understanding which microstructural features cause these high-stress states is crucial for preventing failure. This knowledge is also essential for material design. However, the mechanisms of ductile damage remain poorly understood because several deformation processes occur simultaneously.

The mechanistic complexity underlying these processes arises from multiple coupled phenomena across different length scales. Dislocation glide dominates plastic flow \cite{hirth1983theory,clayton2010nonlinear}. Twinning also contributes to this in low-symmetry systems \cite{murr1997shock, shields1975deformation}. Furthermore, the non-Schmid effects cause dislocation glide, making slip behavior more complex, as it deviates from the classical Schmid law \cite{vitek1970core, duesbery1973effect, vitek2004core,groger2008multiscale}. These underlying complex mechanisms drive stress concentrations near grain boundaries and junctions, which create the conditions for void nucleation \cite{francis2021multimodal,lieberman2016microstructural, gray2014influence}. Thus, understanding the underlying micro-mechanics near grain boundaries is crucial to predicting ductile failure.

Given this mechanistic complexity, various computational approaches have been developed to model and predict ductile damage. Early work by \cite{johnson1981dynamic} introduced a mathematical model for the growth of voids under tensile mean stress, which is applied to spallation
problems through a microscopic to continuous framework. The models were further advanced by incorporating micro-inertial effects, which are relevant to dynamic loading conditions \cite{ortiz1992effect,tong1995inertial,molinari2001micromechanical}.
After recognizing the inherent stochasticity of void nucleation, probabilistic laws were introduced \cite{versino2018computationally, czarnota2008modelling}. In addition, soft-coupled linkage techniques have been used to integrate macroscale damage models and micromechanical calculations to study pore nucleation, as exemplified by \cite{schmelzer2025statistical, bronkhorst2021local}. Bayesian inference and machine-learning techniques have also been introduced for material parameter evaluation and microstructure-sensitive damage prediction \cite{nguyen2021bayesian,kuhn2022identifying,bhamidipati2025bayesian}. In parallel, we recently developed a physics-assisted statistical model that identifies interpretable relationships between microstructural features and stress states with uncertainty quantification \cite{zhang2025physics,dunham2025attribution}, providing the basis for a Bayesian framework targeting extreme stress events.

Despite these advances, several challenges limit our ability to explore damage models and identify damage attributions to microstructural characteristics. First, the available geometry-resolved datasets linking microstructure to void nucleation under dynamical loading are insufficient, since experiments cannot easily capture relevant time scales, and high-fidelity crystal plasticity simulations are computationally intensive \cite{roters2010overview}. These limited data weaken the reliability of results in traditional statistical analyses of extreme events. Second, local stresses near grain boundaries vary widely and show heavy-tailed distributions \cite{schmelzer2025statistical, gehrig2022fft}. Predictive models that emphasize the mean trend, therefore, miss the rare but critical stress concentrations that trigger damage initiation \cite{clayton2010nonlinear, roters2010overview}. Furthermore, modeling polycrystals involves high-dimensional microstructural features. The joint distribution of these high-dimensional microstructural features also shows non-Gaussian characteristics \cite{dunham2025attribution, zhang2025physics}. Consequently, the combination of high dimensionality and non-Gaussianity makes traditional statistical estimation unreliable, requiring rigorous regularization. Although Bayesian analyses are well-suited for uncertainty quantification, previous Bayesian work has focused more on calibrating macroscopic parameters in damage modeling \cite{nguyen2021bayesian, kuhn2022identifying, bhamidipati2025bayesian}. These limitations highlight the need for a probabilistic inference framework that explicitly accounts for extreme events while maintaining physical consistency with material damage.

In this paper, a new Bayesian inference framework is developed to solve the inverse problem of identifying microstructural attributions to extreme stress events. This framework addresses the three fundamental challenges outlined above through an integrated set of key components.
First, we reformulate the variational inference (VI) objective to explicitly prioritize extreme events. In standard VI or sample-based methods, the aim is to optimize over all observations, which tends to emphasize average trends. Our VI method modifies this objective to assign greater weight to high-stress events. We incorporate Extreme Value Theory (EVT) \cite{smith1990extreme, gomes2015extreme} into the likelihood and introduce specialized responsibility weights during the update process. This change reframes rare events not as noise, but as the most informative part of the data. As a result, the inference naturally updates toward microstructural configurations that drive failure, achieving high accuracy in the tails at a lower computational cost.
Second, we construct a hybrid likelihood function that integrates a recently developed physics-based statistical model with extreme value statistics to maintain physical consistency despite data scarcity. Building upon the physics-informed statistical model \cite{dunham2025attribution,zhang2025physics}, which encodes established relationships between microstructural features and local stress responses, we design a likelihood to emphasize tail behavior through EVT-based characterization of stress exceedance probabilities. In this integration, EVT provides the statistical framework for tail behavior, while crystal plasticity mechanics constrains predictions to remain physically realistic. Unlike purely data-driven approaches that risk exploring into nonphysical regimes, or purely mechanistic models that are expensive and lack a way to quantify uncertainty, our hybrid likelihood unites deterministic physical relationships with probabilistic statistical modeling while remaining computationally efficient.
Third, we apply Gaussian Mixture Models (GMMs) \cite{rasmussen1999infinite, huang2017model} in a dimension-reduced latent space to capture the complex and non-Gaussian joint distribution of microstructural features. The large number of microstructural features and their interactions necessitate dimension reduction to make inference tractable. Critically, we represent both prior and posterior distributions as GMMs rather than relying on Gaussian assumptions or Gaussian copulas used in previous Bayesian analyses \cite{nguyen2021bayesian}. Since the underlying microstructural mechanisms, such as different grain orientations, slip system activations, and boundary configurations, cannot be adequately represented by Gaussian distributions, the mixture model provides a flexible way to represent these non-Gaussian statistical properties while remaining analytically tractable for VI updates.

The remainder of this paper is organized as follows. The new Bayesian inference framework, which bridges physics-informed models with extreme-statistics analysis to identify extreme-event attributions, is developed in Section \ref{Sec:Method}. Section \ref{Sec:Data} describes the experimental setup and bicrystal configurations used in this study. We then present validation results in Section \ref{Sec:Results} before concluding in Section \ref{Sec:Discussion} with future research directions.

\section{Methodology}\label{Sec:Method}

\subsection{Problem Formulation and Overall Framework}

Identifying the microstructural features associated with extreme stress events is a crucial inverse problem, and solving it is essential to prevent catastrophic failure and enable safe material design. While forward simulations can predict stress states from given microstructural features, the inverse problem, reasoning from observed extreme events back to their microstructural causes, aims to uniquely identify the attributions. However, unique challenges remain in solving such an inverse problem in the complex material modeling setup.

We formalize this problem as a task of estimating a conditional distribution. For the high-purity polycrystalline metal system, there are microstructural features $\mathbf{x}\in \mathbb{R}^{n}$ that play an essential role in the occurrence of ductile damage, such as elastic
strain and dislocation density. The corresponding stress states near grain boundaries are denoted as $\sigma \in \mathbb{R}^{1}$. Then, extreme events $\mathbf{E} = {\sigma > \bar{\sigma}}$ refer to cases where the stress states exceed a predetermined threshold $\bar{\sigma}$. Here, the goal is to estimate the conditional distribution $P(\mathbf{x}\mid \mathbf{E})$, which captures the microstructural features most likely to lead to extreme events. This conditional distribution not only reveals which microstructural features lead to extreme events but also quantifies the probability of their occurrence.

There are several challenges for directly estimating the conditional distribution $P(\mathbf{x}\mid \mathbf{E})$. First, these extreme damage events are rare. The experimental data contain too few samples for reliable statistics. Second, microstructural features are high-dimensional because polycrystalline systems require representing both individual grains and their interactions. This results in many correlated variables that cannot be ignored, creating a curse of dimensionality in which standard statistical estimators require sample sizes that grow exponentially with dimension. Furthermore, the underlying microstructural mechanisms give rise to non-Gaussian joint distributions that cannot be captured by simplified Gaussian models.

The physics-informed statistical model $f(\mathbf{x})$ developed recently \cite{dunham2025attribution,zhang2025physics}, which describes the relationship between microstructural features and stress, provides a unique tool for efficiently generating information and creating many more samples that overcome the above undersampling difficulty. Yet, such a model alone is still insufficient for accurately estimating the conditional distribution due to uncertainty and model error. A high value of the predicted stress, for instance, does not guarantee an extreme event has actually occurred since the model can only provide probabilistic predictions within a range rather than definitive answers.

\begin{figure}[ht]
    \centering
    \includegraphics[width=1\linewidth]{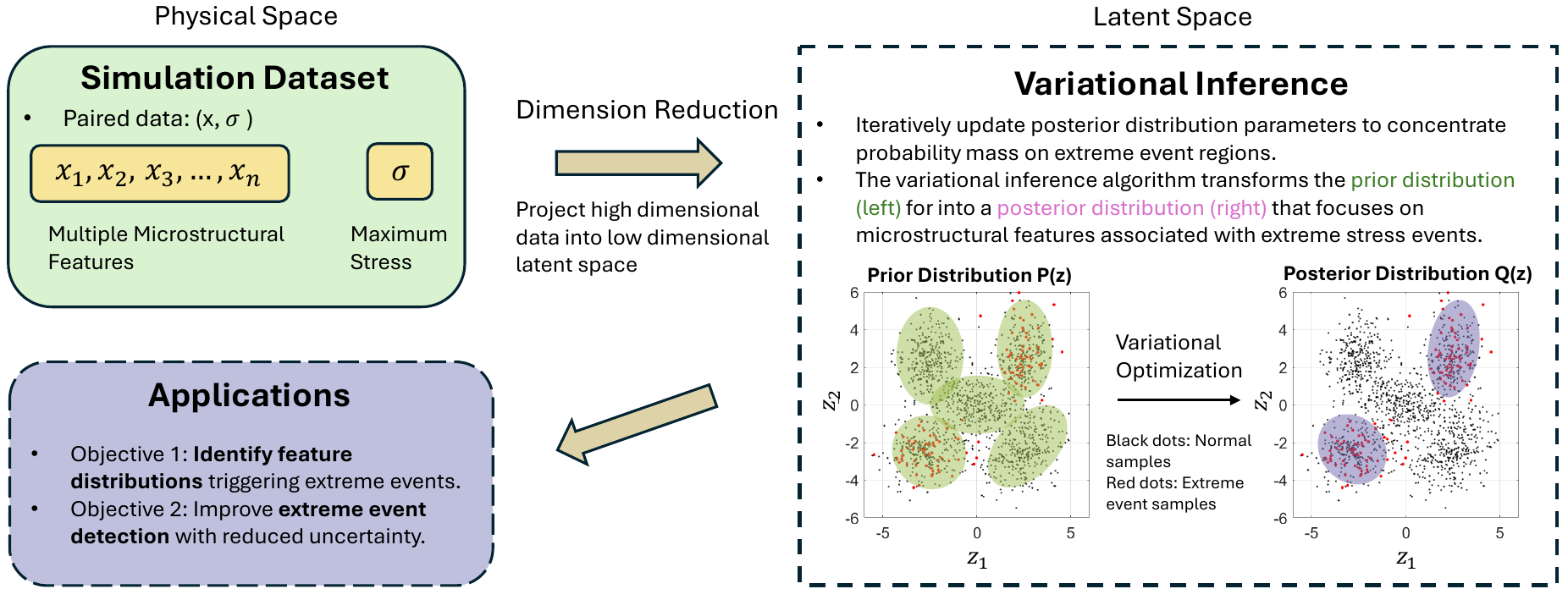}
    \caption{Overview diagram of the general physics-informed variational inference framework.}
    \label{fig:Overview}
\end{figure}

These challenges motivate us to develop a Bayesian inference framework that bridges the physics-informed model and extreme statistics analysis. The former imposes physical constraints, and the latter provides proper uncertainty quantification. As illustrated in Figure \ref{fig:Overview}, the overall framework is summarized into several steps:

\begin{enumerate}
\item To overcome high-dimensionality, the microstructural features $\mathbf{x} \in \mathbb{R}^n$ are first projected into a low-dimensional latent space, becoming the latent variables $\mathbf{z} \in \mathbb{R}^d$, where $d \leq n$. This dimension reduction makes subsequent inference computationally efficient while preserving dominant variability.
\item The joint distributions of latent variables are fitted by GMMs, which is the prior distribution. This flexible representation benefits the capture of non-Gaussian properties arising from the complex underlying dynamics of microstructures.
\item The conditional distribution $P(\mathbf{z} \mid \mathbf{E})$ is then approximated using a variational posterior $Q(\mathbf{z})$, also represented as a GMM, known as the posterior distribution. In the approximation process, the parameters of the posterior distribution are iteratively optimized using a specialized objective, with a likelihood function that combines our physics-informed stress predictions with EVT to emphasize extreme events.
\item The resulting posterior distribution is defined in the latent space, but each sample corresponds to microstructural features in the original physical space. Our dimension-reduction method also allows projection back into the physical space for direct interpretation of extreme-event mechanisms. There are two crucial implications in physical space: (i) identifying the microstructural distributions that trigger extreme stresses, and (ii) improving extreme-event detection with reduced uncertainty.
\end{enumerate}

The details of this Bayesian inference framework are presented in the following subsections.

\subsection{Dimension Reduction}

To overcome the high dimensionality of microstructural features $\mathbf{x}$, dimensional reduction methods are considered to improve computational efficiency. Although nonlinear dimension-reduction methods such as autoencoders and manifold learning \cite{carreira1997review,van2009dimensionality,mai2020finding} offer alternatives, in this work, we use Principal Component Analysis (PCA) \cite{wold1987principal, abdi2010principal, jolliffe2016principal} for its simplicity, interpretability, and ability to map results back to the original feature space. As a linear method, PCA retains dominant modes of variability while avoiding overfitting in limited datasets.

Specifically, we project microstructural features $\mathbf{x}$ into a latent space yielding $\mathbf{z}\in \mathbb{R}^d$ with $d \leq n$. Thus, the latent variables $\mathbf{z}$, which are also called principal components (PCs) in PCA cases, serve as a surrogate of microstructural states. The relationships between the PCs and microstructural features are discussed in Appendix \ref{Sec:PCA}. Working in the latent space $P(\mathbf{z}\mid \mathbf{E})$ instead of the original space $P(\mathbf{x}\mid \mathbf{E})$ makes subsequent VI tractable. Results can then be projected back to the physical space for interpretation.

\subsection{Bayesian Updating Framework}

As discussed above, the predictive model alone is insufficient for estimating the conditional distribution due to inherent uncertainties and model error, the Bayesian framework is introduced to solve the inverse problem. This principled framework combines our prior knowledge of microstructural features with the observations of extreme stress events, allowing us to formally quantify uncertainty in the inverse problem. Our goal is to compute the posterior distribution $P(\mathbf{z}|\mathbf{E})$, which represents our updated probability quantification about latent microstructural states given that an extreme event has occurred.

Following Bayes' theorem, the posterior is given by:
\begin{equation}\label{Eq:Bayesian}
    P(\mathbf{z}\mid\mathbf{E}) = \frac{P(\mathbf{E}\mid\mathbf{z})P(\mathbf{z})}{P(\mathbf{E})},
\end{equation}
where $P(\mathbf{E}\mid\mathbf{z})$ represents the likelihood and $P(\mathbf{z})$ is the prior distribution describing latent factors for the entire dataset.
Direct computation of the conditional distribution is analytically intractable because the denominator in Equation \eqref{Eq:Bayesian} cannot be evaluated in closed form: $
P(\mathbf{E})  =  \int P(\mathbf{E}\mid \mathbf{z})\,P(\mathbf{z}) \text{d}\mathbf{z}.
$

To address this intractability, approximation methods are required. Among these methods, VI provides a computationally efficient strategy for identifying the cause of extreme events, which transforms the integration problem into a tractable optimization problem \cite{blei2017variational}. The framework has several advantages over alternative sampling-based methods, such as Markov Chain Monte Carlo (MCMC). First, MCMC methods rely on sampling the posterior \cite{neal2011mcmc,robert1999monte}, which becomes inefficient when the target conditional distribution is concentrated in small regions of the feature space. Accurately recovering posterior distribution requires a long Markov chain. In contrast, VI directly optimizes the approximation to the posterior. Thus, a proper parameter update algorithm will make the approximation efficient in the extreme event regime.
Second, our goal is not only to recover the posterior but also to characterize its structure in the tails. This allows us to explicitly modify the optimization objective to focus on tail behavior, for instance, by integrating EVT into the likelihood.
Third, MCMC diagnostics are often unreliable for checking convergence in the tails of a distribution. In contrast, VI offers a deterministic and monotonic convergence guarantee by maximizing the objective function, which is here referred to as the Evidence Lower Bound (ELBO) \cite{jordan1999introduction,blei2017variational}. The ELBO also serves as a principled and computationally tractable measure of approximation \cite{bishop2006pattern}.

We utilize a parameterized distribution $Q(\mathbf{z};\theta)$ to approximate a conditional distribution of latent variables. The goal is to minimize the difference between $Q(\mathbf{z};\theta)$ and the exact posterior distribution,
\begin{equation}
Q^*(\mathbf{z};\theta) =\underset{Q(\mathbf{z};\theta) \in \mathscr{Q}}{\arg \min} \mathrm{KL}(Q(\mathbf{z};\theta) \| P(\mathbf{z}\mid\mathbf{E}))
\end{equation}
\begin{equation}
\begin{aligned}\label{eq:KL}
\mathrm{KL}\left(Q(\mathbf{z};\theta) \|
P(\mathbf{z}\mid\mathbf{E})\right),
& = \int Q(\mathbf{z};\theta)\,\log\frac{Q(\mathbf{z};\theta)}{P(\mathbf{z}\mid\mathbf{E})}\,d\mathbf{z}\\
& = \mathrm{KL}\left(Q(\mathbf{z};\theta) \|
P(\mathbf{z})\right) - \mathbb{E}_Q\left[\log P(\mathbf{E}\mid \mathbf{z})\right] + \mathbb{E}_Q\left[P(\mathbf{E})\right],
\end{aligned}
\end{equation}
where $\mathscr{Q}$ represents a family of densities over the latent feature space, $Q^*(\mathbf{z};\theta)$ is the optimal approximation of the conditional distribution, and $\mathrm{KL}$ stands for the Kullback–Leibler divergence, which is an information measurement to quantify the difference between two distributions \cite{kleeman2002measuring, hershey2007approximating, majda2018model}. To avoid computation of the normalization constant $P(\mathbf{E})$, we optimize an alternative objective that dropped this constant term in Equation \eqref{eq:KL}:
\begin{equation}
    \mathrm{ELBO}(Q)=\mathbb{E}_Q[\log P(\mathbf{E} \mid \mathbf{z})]-\mathrm{KL}(Q(\mathbf{z};\theta) \| P(\mathbf{z})) .
\end{equation}
Maximizing the ELBO is equivalent to minimizing the KL divergence in Equation \eqref{eq:KL}. While standard VI computes likelihoods over all observations, our formulation restricts the likelihood to extreme events only. This modification has two key effects. The first term $\mathbb{E}_Q[\log P(\mathbf{E}\mid \mathbf{z})]$ places more weight on extreme events, which naturally pushes the variational distribution toward regions of the latent space where microstructural configurations are most likely to produce stress exceedance. The second term penalizes deviations from the prior distribution, which is especially important for preventing overfitting when data in the tail regions are limited. Together, these two terms preserve the mathematical structure of standard VI while systematically emphasizing extreme event information that would otherwise receive insufficient attention in conventional likelihood-based objectives.

\subsection{Physics-Informed Stress Likelihood Model}\label{subsec:likelihood}

Having established the ELBO formulation, we now specify how the likelihood term
$\mathbb{E}_Q[\log P(\mathbf{E}\mid \mathbf{z})]$ is constructed in the context of extreme stress events.
A key difficulty is that stress exceedance, while critical for failure, is rare in the data and therefore poorly captured by standard likelihood formulations.
Relying solely on empirical stress exceedance frequencies would underestimate tail behavior and amplify model uncertainty.
Thus, we introduce a physics-informed likelihood based on extreme value theory, which provides a principled way to extrapolate beyond the limited observed extremes.

Two data regimes are considered for likelihood estimation. When extreme stress observations $\sigma_i$ are available, the likelihood can be defined through the exceedance indicator $e_i = \mathbf{1}\{\sigma_i > \bar{\sigma}\}$, with $\bar{\sigma}$ a chosen threshold.  When direct stress is unavailable, we instead evaluate a surrogate stress $\tilde{\sigma}(\hat{\mathbf{x}})$, where $\hat{\mathbf{x}}$ is the reconstructed microstructural feature vector from the latent variable $\mathbf{z}$.
In this regime, the probability of exceedance $P(E \mid \mathbf{z})$ is determined by fitting the tail of $\tilde{\sigma}(\hat{\mathbf{x}})$ with an EVT distribution.

Concretely, we model the tail using a heavy-tailed Fr\'{e}chet distribution \cite{ramos2020frechet}. Here, the Fr\'{e}chet family naturally captures the heavy-tailed behavior observed in polycrystalline stress distributions \cite{schmelzer2025statistical}, so it is well-suited for modeling stress extremes,
\begin{equation}
    p(y ; s, \alpha,m) = \frac{\alpha}{s}\left(\frac{y-m}{s}\right)^{-\alpha-1}
    \exp\left[-\left(\frac{y-m}{s}\right)^{-\alpha}\right], \quad y>0,
\end{equation}
where $y = \tilde{\sigma}(\hat{\mathbf{x}}) - \bar{\sigma}$, $s>0$ is a scale parameter, $m$ is a local parameter, and $\alpha>0$ controls tail heaviness.
The parameters $(s, \alpha,m)$ are estimated via maximum likelihood estimation (MLE) \cite{pan2002maximum}.

Consequently, our framework does not only rely on a single likelihood formulation. Instead, it combines two sources of information about extreme events: direct observations when available, and physics-informed predictions from the EVT-tail model when data are sparse. This combination ensures the likelihood term $\mathbb{E}_Q[\log P(\mathbf{E} \mid \mathbf{z})]$ emphasizes tail behavior. As a result, VI concentrates the posterior $Q(\mathbf{z})$ in regions of latent space associated with high-stress configurations rather than average patterns.

\subsection{Non-Gaussian Prior and Posterior Representation}

In the latent space, the prior distribution $P(\mathbf{z})$ describes the variability of latent variables associated with the observed microstructural features, independent of any extreme event information. Since PCA is a linear transformation, the empirical distribution of $\mathbf{z}$ generally remains non-Gaussian, reflecting the heterogeneity of the underlying microstructure. To capture these characteristics, we represent the prior with a Gaussian mixture model (GMM),
\begin{equation}
    P (\mathbf{z})=\sum_{k=1}^K \omega_k \mathcal{N}\left(\mathbf{z} \mid \nu_k, \Lambda_k\right),
\end{equation}
where $\omega_k$ are mixture weights and $(\nu_k,\Lambda_k)$ are the mean and covariance of component $k$, fitted to the projected features $\mathbf{z}$ by Expectation-Maximum (EM) algorithm. The number of mixture components $K$ can be determined using the Bayesian Information Criterion (BIC) \cite{neath2012bayesian} or Akaike Information Criterion (AIC) \cite{akaike2025akaike}.

The variational approximation adopts the same mixture family,
\begin{equation}
Q(\mathbf{z};\theta)=\sum_{k=1}^{K}\pi_k\,\mathcal{N}\left(\mathbf{z}\mid \mu_k,\Sigma_k\right),
\end{equation}
with $\theta=\{\pi_k,\mu_k,\Sigma_k\}_{k=1}^K$ (the number of components $K$ is same as prior distribution) optimized by maximizing the ELBO from the previous subsection. This choice provides a flexible, explicitly non-Gaussian variational family $\mathscr{Q}$ for $P(\mathbf{z}\mid \mathbf{E})$.

The use of GMMs is motivated by both flexibility and practicality. In polycrystalline materials, the latent space typically exhibits a complex multimodal structure because different competing microstructural mechanisms create distinct patterns in the data. Alternative approaches, such as Gaussian copulas \cite{song2009joint} or Nataf transformations \cite{lebrun2009innovating}, assume Gaussian distributions that work well for certain cases but struggle with the stronger non-Gaussian behavior we observe here. In contrast, GMMs can approximate a wide range of non-Gaussian distributions with adjustable complexity while maintaining the analytical tractability needed for efficient VI updates. An important consideration is that PCA, being a linear transformation, preserves the non-Gaussian distributional characteristics from the original feature space in the latent representation. This property makes GMMs particularly well-suited for modeling both the prior and posterior distributions in our framework. The approach ultimately provides a balance between model interpretability and the representational flexibility required to capture the diverse microstructural mechanisms that drive extreme stress events.

\subsection{Extreme Event Focused Variational Inference Updates}

The posterior approximation is updated through an iterative scheme analogous to the EM algorithm, but modified to incorporate extreme-event likelihood information.

In each iteration, we first compute the responsibilities $r_{ik}$, which represent the soft assignment of each extreme event data $z_i$ to component $k$. The responsibilities are obtained by maximizing the ELBO with respect to the soft assignments with details given in Appendix \ref{appendix:Derivation}:
\begin{equation}
    \tilde r_{ik} = \frac{\omega_k \mathcal{N}\left(\mathbf{z}_i \mid \nu_k, \Lambda_k\right) P\left(\sigma_i>\bar{\sigma} \mid \mathbf{z}_i\right)}{\pi_k \mathcal{N}\left(\mathbf{z}_i \mid \mu_k, \Sigma_k\right)},
\end{equation}
\begin{equation}
\begin{aligned}
\log \tilde{r}_{i, k} & =\log \omega_k-\frac{1}{2} \log \left|\boldsymbol{\Lambda}_k\right|-\frac{1}{2}\left(\mathbf{z}_i-\boldsymbol{\nu}_k\right)^T \boldsymbol{\Lambda}_k^{-1}\left(\mathbf{z}_i-\boldsymbol{\nu}_k\right), \\
& +\log P\left(\sigma_i>\bar{\sigma} \mid \mathbf{z}_i\right)-\left[\log \pi_k-\frac{1}{2} \log \left|\boldsymbol{\Sigma}_k\right|-\frac{1}{2}\left(\mathbf{z}_i-\boldsymbol{\mu}_k\right)^T \boldsymbol{\Sigma}_k^{-1}\left(\mathbf{z}_i-\boldsymbol{\mu}_k\right)\right],
\end{aligned}
\end{equation}
and the normalized responsibilities are
\begin{equation}
    r_{ik} = \frac{\tilde r_{ik}}{\sum_{j=1}^K \tilde r_{ij}},
\end{equation}
where the likelihood term $P(\sigma_i>\bar{\sigma} \mid \mathbf{z}_i)$ emphasizes samples associated with stress exceedances, as described in Section \ref{subsec:likelihood}. Unlike conventional mixture updates, these responsibilities are weighted not only by the prior density but also by the probability of stress exceedance, thereby giving influence to factors associated with rare but critical events.

The parameters of the posterior mixture are then updated by taking weighted averages of the current responsibilities:
\begin{equation}
\pi_k=\frac{1}{N_E} \sum_i r_{i k}, \quad \mu_k=\frac{\sum_i r_{i k} \mathbf{z}_i}{\sum_i r_{i k}}, \quad \Sigma_k=\frac{\sum_i r_{i k}\left(\mathbf{z}_i-\mu_k\right)\left(\mathbf{z}_i-\mu_k\right)^{\top}}{\sum_i r_{i k}} .
\end{equation}
Here, $N_E$ denotes the effective number of extreme-event weighted samples, accounting for observed extreme stress events or samples with high predicted extreme-event likelihood from the statistical model. This normalization ensures posterior weights $\pi_k$ account for both how frequently a component appears in the data and how strongly it is associated with extreme stress events. We repeat this two-step process until the ELBO converges. This EM-like structure keeps computation tractable while systematically incorporating extreme-event information into posterior updates. This balances prior knowledge with the focus on rare but critical tail events.

This update procedure is formalized in Algorithm \ref{alg:extreme_vi}. The algorithm alternates between computing responsibilities that emphasize extreme events and updating posterior parameters based on these weighted assignments until convergence.

\begin{algorithm}[htbp]
\caption{Extreme-Event-Focused Variational Inference Updates}
\label{alg:extreme_vi}
\begin{algorithmic}[1]
\State \textbf{Input:} data $\{\mathbf{z}_i\}_{i=1}^N$, initial parameters start from prior distribution $(\pi_k, \mu_k, \Sigma_k)$
\Repeat
    \State \textbf{First Step: Responsibility computation}
    \For{$i = 1,\dots,N$}
        \For{$k = 1,\dots,K$}
            \State Compute unnormalized responsibility:
            \[
            \tilde r_{ik} \leftarrow
            \frac{\omega_k \, \mathcal{N}(\mathbf{z}_i \mid \nu_k,\Lambda_k)\,
            P(\sigma_i>\bar{\sigma}\mid \mathbf{z}_i)}
            {\pi_k \, \mathcal{N}(\mathbf{z}_i \mid \mu_k,\Sigma_k)}
            \]
        \EndFor
        \State Normalize: $r_{ik} \leftarrow \tilde r_{ik} / \sum_{j=1}^K \tilde r_{ij}$
    \EndFor
    \State \textbf{Second Step: Parameter updates}
    \For{$k = 1,\dots,K$}
        \State Update mixture weight:
        \[
        \pi_k \leftarrow \frac{1}{N_E}\sum_i r_{ik}
        \]
        \State Update mean:
        \[
        \mu_k \leftarrow \frac{\sum_i r_{ik}\mathbf{z}_i}{\sum_i r_{ik}}
        \]
        \State Update covariance:
        \[
        \Sigma_k \leftarrow \frac{\sum_i r_{ik}(\mathbf{z}_i-\mu_k)(\mathbf{z}_i-\mu_k)^{\top}}{\sum_i r_{ik}}
        \]
    \EndFor
\Until{convergence}
\State \textbf{Output:} posterior parameters $(\pi_k, \mu_k, \Sigma_k)$
\end{algorithmic}
\end{algorithm}

\section{Experimental Setting and Data Availability}\label{Sec:Data}

\subsection{Dataset Description and Configuration}

The dataset comes from the crystal plasticity simulations \cite{dunham2025attribution} of two bicrystal configurations: one with the grain boundary plane perpendicular to the direction of compressive loading and one with the grain boundary plane parallel to the direction of loading. In these bicrystal configurations, we fix the microstructure and vary the initial crystallographic orientation of each grain, then apply loading conditions typical of the nucleation regime of damage \cite{bronkhorst2021local, jones2018stress,versino2018computationally}. The maximum stress states in a cylinder are obtained by compiling the results from each set of calculations. For each bicrystal configuration, 800 simulations are performed. Among these, 546 samples (perpendicular case) and 617 samples (parallel case) exhibit their maximum stress values located near the grain boundary, and these constitute the dataset used in our subsequent analysis. To define extreme events, we set the stress threshold $\bar{\sigma}$ such that the upper 5\% of stress realizations are classified as exceedances.
The threshold of 5\% aligns with the physical hypothesis that, spatially, damage nucleation events are extreme-event processes driven by localized stress events near weak atomistic defects. Recently, the authors in \cite{schmelzer2025statistical} developed a void nucleation criterion based on the spatial appearance frequencies of both polycrystalline stress distributions and grain boundary nucleation strength as assessed by molecular dynamics calculations.

\subsection{Statistical Model for Stress at Grain Boundary}

Building upon previous work \cite{dunham2025attribution,zhang2025physics} that established statistical relationships between microstructural features and stress states, we consider several important microstructure features.

A set of microstructural features is extracted to capture the mechanical response and crystallographic attributes of each grain. First, the elastic stiffness tensor is rotated into the global frame, giving components $\mathcal{C}_{ij,Gn}$. Here, $\mathcal{C}_{i j, G n}$ denotes the $i j$-th entry of a $6 \times 6$ matrix in Voigt notation, where $i,j$ represent the row and column indices. Similarly, grain-averaged elastic strain $\mathbf{E}_{ij,Gn}^e$ is included to capture the mean deformation state within each grain. Additionally, the rate of plastic deformation is characterized by the eigenvalues $\lambda_{i, G n}$ and eigenvectors $\mathbf{v}_{i, G n}$, with $i= 1,2,3$, of the plastic velocity gradient. Comparisons are made either between corresponding principal directions across a boundary, i.e. $\lambda_{i,Gn} \lambda_{i,Gm}$, or between hotspot values, i.e. ${\lambda_{i,G n}}_L$ and ${\mathbf{v}_{i, G n}}_L$. Here, the subscript $L$ indicates that the quantity is measured using microstructural information local to the elevated stress state within the grain. Third, non-Schmid factors are included to represent slip system interactions beyond the classical Schmid law. Although each grain has 48 such factors, we retain only the top five after ranking them in descending order of magnitude, since any arbitrary deformation may be accommodated by five independent slip systems \cite{taylor_mechanism_1934}. The non-Schmid factors evaluated using the local stress state, $\{\hat{\tau}_{G n}\}_L$, are also included. Finally, the statistically stored dislocation density $\sqrt{\rho_{ssd}}$ is used as a feature to capture dislocation-based hardening.
All features are extracted at the integration point where the von Mises stress reaches its maximum, and paired with corresponding quantities from the adjacent grain across the boundary.

The microstructural descriptors introduced above can be systematically linked to the maximum stress near grain boundaries through a quadratic regression model \cite{dunham2025attribution,zhang2025physics}. The form of the model is as following:

\begin{equation}\label{Eq:stress_function}
\begin{aligned}
\sigma_{\text {model }} & =\beta_0 \sqrt{\rho_{\mathrm{ssd}}}+\sum_n^{N_{\mathrm{gr}}} \sum_{i=1}^3 \beta_{1 n i} \lambda_{i, G n}+\sum_n^{N_{\mathrm{gr}}} \sum_{i=1}^3 \beta_{2 n i} \lambda_{i, G n}^{\max } \\
& +\sum_n^{N_{\mathrm{gr}}} \sum_{i, j=1}^3 \beta_{3 n i j} \mathbf{E}_{i j, G n}^e+\sum_n^{N_{\mathrm{gr}}} \sum_{i=1}^3 \beta_{4 n i} \mathcal{C}_{i i, G n} \\
& +\sum_{m>n}^{N_{\mathrm{gr}}} \sum_{i, j=1}^3 \beta_{5 n m i j} \mathbf{E}_{i j, G n}^e \mathbf{E}_{i j, G m}^e+\sum_{m>n}^{N_{\mathrm{gr}}} \sum_{i=1}^3 \beta_{6 n m i} \mathcal{C}_{i i, G n} \mathcal{C}_{i i, G m} \\
& +\sum_n^{N_{\mathrm{gr}}} \sum_{i, j=1}^3 \beta_{7 n i j}\left(\mathbf{E}_{i j, G n}^e\right)^2+\sum_n^{N_{\mathrm{gr}}} \sum_{i=1}^3 \beta_{8 n i j}\left(\mathcal{C}_{i i, G n}^e\right)^2 \\
& +\sum_{m>n}^{N_{\mathrm{gr}}} \sum_{i, j=1}^5 \beta_{9 n m i j} \hat{\tau}_{i, G n} \hat{\tau}_{j, G m}+\sum_{m>n}^{N_{\mathrm{gr}}} \sum_{i=1}^5 \beta_{10 n m i} \hat{\tau}_{i, G n}^{\max } \hat{\tau}_{i, G m}^{\max } \\
& +\sum_n^{N_{\mathrm{gr}}} \sum_{i=1}^3 \beta_{11 n i}\left(\mathbf{v}_{i, G n} \cdot \mathbf{v}_{i, G m}\right)^2 + \sum_n^{N_{\mathrm{gr}}} \sum_{i=1}^3 \beta_{12 n i}\left(\mathbf{v}_{i, G n}^{\max } \cdot \mathbf{v}_{i, G m}^{\max }\right)^2,
\end{aligned}
\end{equation}
where $\sigma_{\text{model}}$ denotes the predicted maximum stress. The coefficients $\beta$ are regression parameters learned from simulation data, $N_{\mathrm{gr}}$ is the number of grains, and the feature notation follows the definitions in the previous subsection.

\subsection{Computational Experiment Data Settings}

The number of bicrystal simulations available for analysis is on the order of a few hundred, which is small relative to the dimensionality of the microstructural feature space. Moreover, extreme stress events are rare, leading to an imbalanced dataset with limited tail information. These restrictions make it impractical to rely solely on direct simulation data to validate the proposed inference framework. Therefore, the statistical model plays a crucial role in the inference.

In the following, we conduct two types of validation experiments.
First, we implement perfect model tests based only on the statistical model, in which the true functional form of the stress response is known. Five thousand synthetic microstructural features are generated by sampling from the fitted GMM prior distribution, then computing their corresponding stress states using our established statistical stress function shown in Equation \eqref{Eq:stress_function}. This synthetic dataset is split into 4000 training and 1000 test samples. These tests provide a controlled setting to isolate the performance of the VI methodology and exclude the influence of model error.
Second, we perform mixed model-data simulation tests, augmenting the limited bicrystal data with synthetic samples drawn from the fitted prior distribution. We generate 1400 synthetic samples from the prior distribution and combine 700 of these with 300 experimental bicrystal simulation samples to form our training set (1000 samples total). The remaining 700 synthetic samples are combined with the remaining experimental data to create the test set. The mixed model-data experiment provides an effective dataset that preserves the statistical structure of the observations and enables a more realistic assessment of the predictive capability of the framework under realistic conditions.

\section{Computational Experiment Results and Analysis}\label{Sec:Results}

\subsection{Perfect Model Test Results}

To validate the capability of our VI method for targeted conditional distribution recovery and extreme event detection, we first conduct perfect model tests in which synthetic microstructural feature data are generated and the corresponding stress state is computed using our established stress function \eqref{Eq:stress_function}. Under this controlled condition, we evaluate three distinct methods as follows:

\begin{enumerate}
    \item GMM-VI Method (Proposed): As depicted in Figure \ref{fig:Overview}, the Gaussian mixture variational inference approach in this study aims to closely match the target conditional distribution of the feature space. We iteratively refine the parameters of the Gaussian components to maximize the ELBO. Additionally, we employ a responsibility-weighting method to highlight the importance of extreme-event likelihoods in the analysis.
    \item MCMC: MCMC is a sampling-based inference strategy, which iteratively proposes candidate states in the latent space and accepts or rejects them according to the probability ratio \cite{andrieu2008tutorial}. These samples, in principle, asymptotically follow the exact posterior distribution. However, MCMC is extremely computationally expensive when the latent dimensionality is high, which limits its practicality compared to VI.
    \item Empirical Distribution: This distribution is obtained by directly fitting the GMM to observed feature space according to extreme events in the latent space. It represents a non-Bayesian baseline that captures the conditional distribution based solely on observations.
\end{enumerate}

\begin{figure}[ht]
    \centering
    \includegraphics[width=1\linewidth]{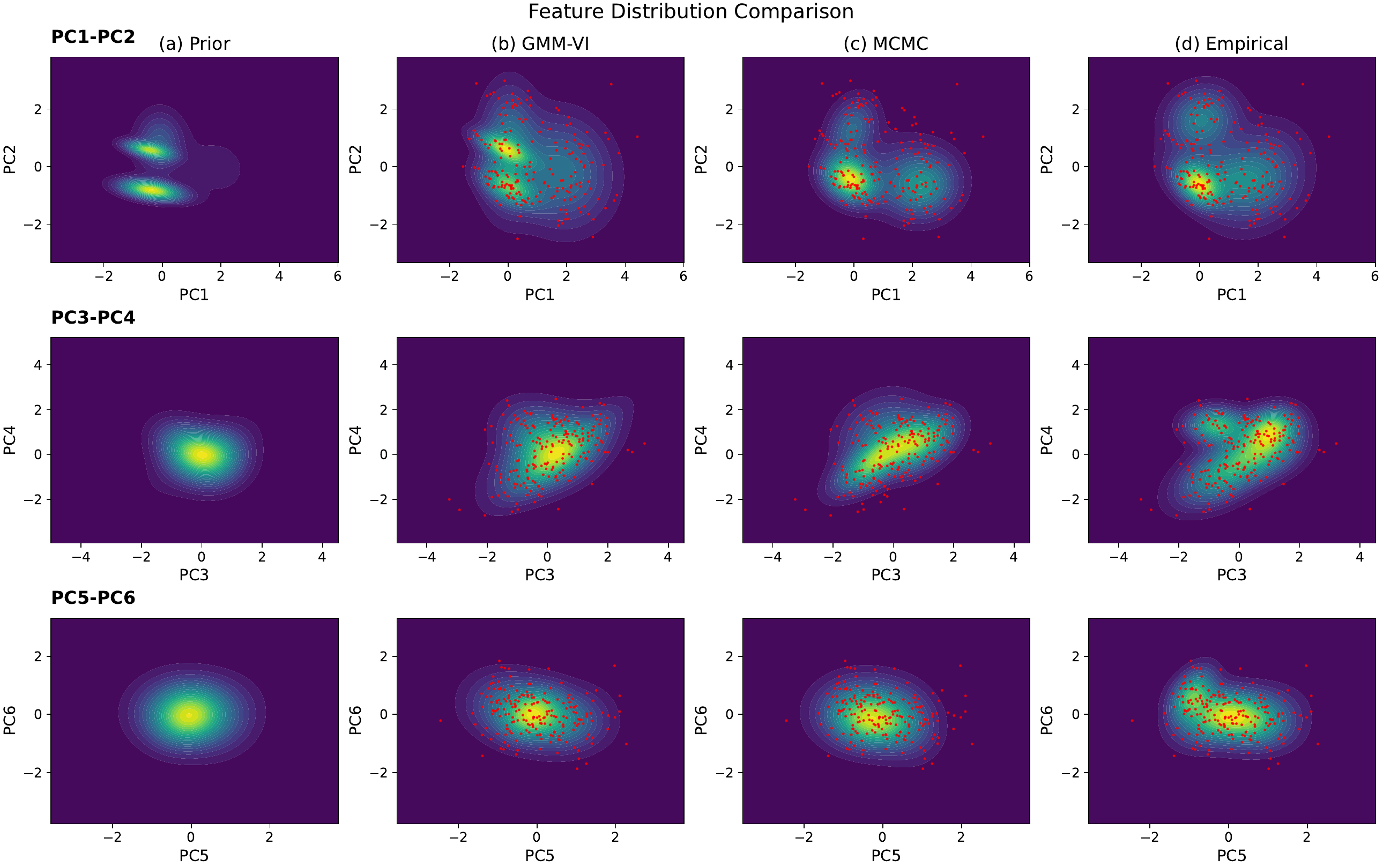}
    \caption{Comparison of posterior distributions in PCA latent space across different inference methods. The figure shows PDF contours and scatter plots for three pairs of principal components. (a) Prior GMM distribution fitted to all training data, (b) Variational inference (GMM-VI) posterior targeting extreme events, (c) MCMC posterior samples for the same target distribution, and (d) Empirical GMM fitted directly to observed extreme events. Colored density maps (purple to yellow) indicate probability density from low to high. Red scatter points show the locations of extreme events.}
    \label{fig:pdf_compare_multiple_pc_pairs}
\end{figure}

These three methods are tested across two key aspects: posterior distribution recovery and extreme event classification. Figure \ref{fig:pdf_compare_multiple_pc_pairs} compares posterior (or estimated) distributions recovered by each method. Each row corresponds to a different pair of principal components (PCs), providing a comprehensive view of the posterior structure in the most dominant dimensions of the PCA latent space. Panel (\ref{fig:pdf_compare_multiple_pc_pairs}a) shows the prior distribution of all components in latent space, which is directly fitted by all training data (bicrystal case under perpendicular grain boundary) by GMM. The number of prior Gaussian components is estimated to be four by the BIC as shown in Panel (\ref{fig:bic_analysis}a) of Figure \ref{fig:bic_analysis}. The prior distribution serves as the initial state for variational inference (VI) and provides the baseline distribution for MCMC candidate state generation.
Panel (\ref{fig:pdf_compare_multiple_pc_pairs}b) demonstrates that the VI posterior effectively concentrates the distribution in regions associated with high stress exceedance, showing clear adaptation from the broad prior to a focused posterior that highlights extreme events.
Panel (\ref{fig:pdf_compare_multiple_pc_pairs}c) presents the estimated distribution for features by the MCMC method. While the MCMC method can, in principle, recover the exact posterior distribution, it struggles in practice when extreme events are rare. At each MCMC sampling step, the algorithm proposes new candidate samples by sampling from the broad prior distribution, but extreme events occupy only a small fraction of this space. As a result, most proposals miss the relevant regions entirely, leading to low acceptance rates and wasted computation. Even after many iterations, large portions of the posterior remain unexplored. Under the same computational budget, MCMC simply cannot match the efficiency of our directed variational approach, which systematically guides the search toward extreme-event regions rather than wandering through the whole feature space.
Panel (\ref{fig:pdf_compare_multiple_pc_pairs}d) shows the empirical distribution obtained by fitting a GMM directly to the observed extreme events. This approach relies solely on the limited labeled extreme events in the training data, without accounting for physics-informed likelihoods or prior information, and thus serves as a data-driven baseline for comparison. This estimation is less statistically robust, as small changes in the training sample could significantly alter the fitted distribution. In contrast, the VI posterior is stabilized by combining the physics-informed likelihood with the prior distribution, yielding smoother contours that generalize better beyond the specific observed extremes.

\begin{figure}[ht]
    \centering
    \includegraphics[width=1\linewidth]{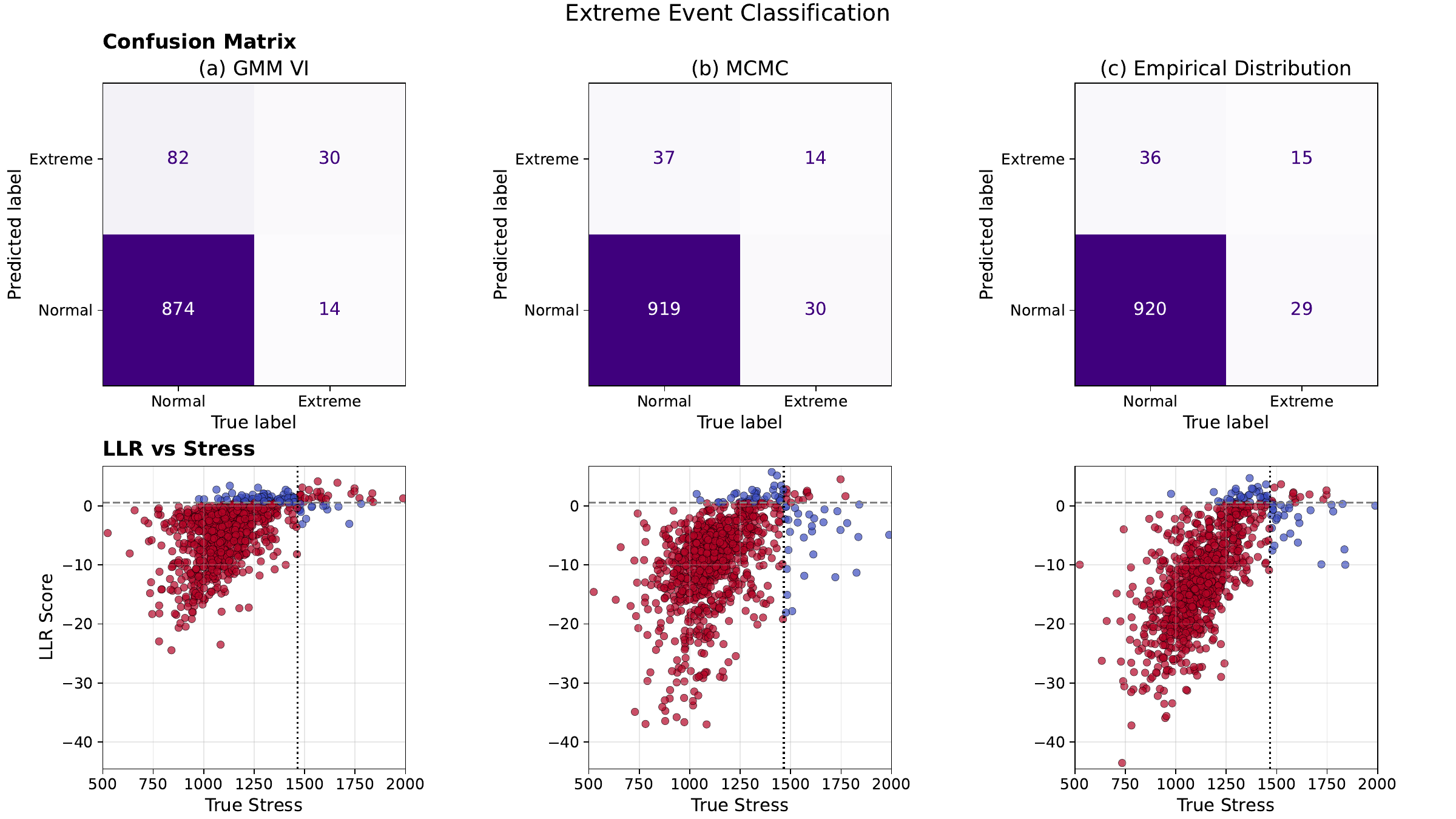}
    \caption{Performance evaluation of extreme event classification using three different inference methods: (a) GMM-VI posterior, (b) MCMC posterior, and (c) Empirical distribution estimation. Top row shows confusion matrices for binary classification (Normal vs. Extreme). Numbers in cells represent counts of true positives, false positives, true negatives, and false negatives. Bottom row displays LLR scores plotted against true stress values for each method, where points are colored by prediction correctness (red = correct, blue = incorrect). The vertical dashed line indicates the stress threshold $S_{th}$, while the horizontal line shows the LLR decision threshold.}
    \label{fig:classification_compare}
\end{figure}

Beyond distribution recovery, another measurement of performance is their extreme event classification performance as presented in Figure \ref{fig:classification_compare}. To measure the performance of event classification, the log-likelihood ratio (LLR) is introduced to understand how much more likely a certain sample $\mathbf{z}'$ is to result in extreme events. The LLR of a certain sample is computed as:
\begin{equation}
    \text{LLR}(\mathbf{z}') = \log q(\mathbf{z}')-\log p(\mathbf{z}')
\end{equation}
where $q(\cdot)$ represents the estimated posterior distribution (from VI, MCMC, or empirical fitting) and $p(\cdot)$ is the prior distribution. A greater result indicates that a sample is more likely to be associated with extreme events, while a smaller result suggests the sample is more characteristic of the normal events. This approach exploits the distributional shift between normal conditions (prior) and extreme conditions (posterior). Unlike basic stress thresholding, the LLR approach takes into account the complete uncertainty associated with both the prior and posterior distributions, which adds additional uncertainty quantification not provided by a thresholding approach. By considering how these two distributions overlap, the LLR helps minimize the chances of misclassifying normal events as extreme, which results in a more trustworthy classification process overall.

Figure \ref{fig:classification_compare} displays the classification performance for three methods. While LLR$> 0$ provides a natural threshold (posterior exceeds prior), we adopt a slightly higher threshold of 0.5 to keep the proportion of predicted extreme events manageable for practical analysis and validation.
The top row presents confusion matrices for binary classification, where each cell indicates the number of predictions in each category based on the LLR detection.
Panel (\ref{fig:classification_compare}a) shows that the VI method captures the largest number of extreme events while incurring the fewest false negatives. This high sensitivity is especially critical in materials applications, since missing extreme events can lead to catastrophic failures. The trade-off, however, is that VI also produces more false positives than MCMC and the empirical approach, as shown in Panels (\ref{fig:classification_compare}b) and (\ref{fig:classification_compare}c). It reflects the balance between detecting critical extremes and avoiding overly conservative predictions that misclassify some normal cases.
The bottom row displays LLR scores plotted against stress values for the test data, with points colored according to their prediction correctness. The red color represents correctness, while the blue color indicates incorrectness. These scatter plots reveal additional insights beyond the binary classification metrics. It is evident that, although the VI method yields some false negatives, all are close to the LLR threshold, suggesting marginal cases rather than clear misses of obvious extreme events. In contrast, MCMC and the empirical methods show more scattered false negatives across different LLR ranges, reflecting less consistent classification performance.

\begin{figure}[ht]
    \centering
    \includegraphics[width=1\linewidth]{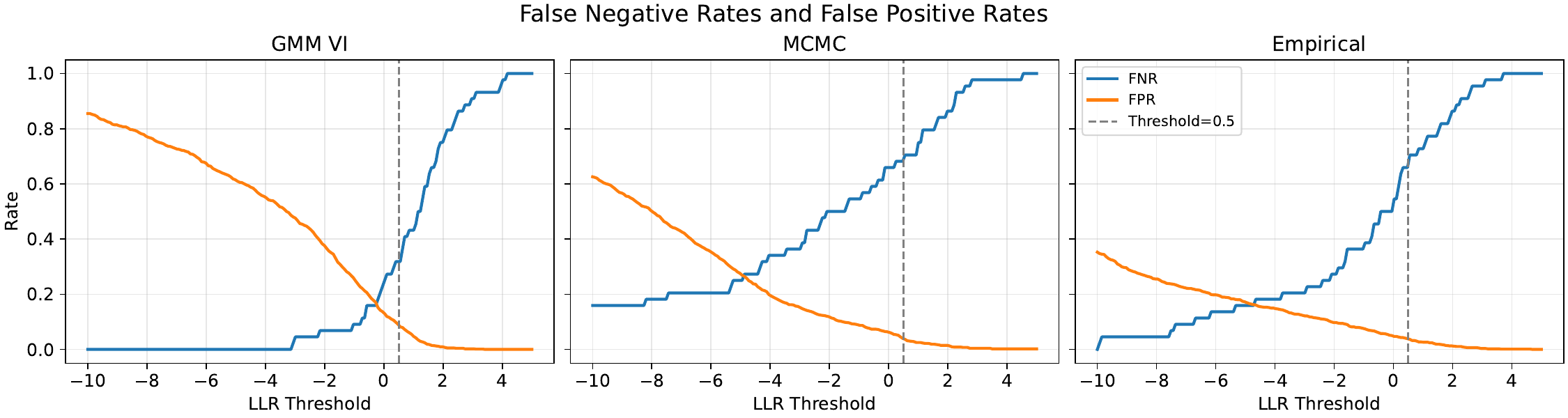}
    \caption{False negative rate and false positive rate as functions of LLR threshold for three methods in the perfect model test.}
    \label{fig:FNRandLLR}
\end{figure}

To provide a broader context for threshold selection, we evaluate performance across a range of values using labeled validation data. We quantify performance using false negative rate (FNR, proportion of missed extremes) and false positive rate (FPR, proportion of false alarms):
\begin{equation} \text{FNR} = \frac{\text{FN}}{\text{FN} + \text{TP}}, \quad \text{FPR} = \frac{\text{FP}}{\text{FP} + \text{TN}},
\end{equation}
where FN and FP stand for false negative and false positive, respectively. 
Figure \ref{fig:FNRandLLR}  shows how FNR and FPR vary with the LLR threshold. The pattern is intuitive: set a higher threshold, and fewer extreme events (FNR rises) will be caught, but fewer false positives (FPR falls) will be generated. Importantly, GMM-VI consistently misses fewer extreme events than the other methods across nearly the entire threshold range, suggesting better calibration.

\subsection{Mixed Model-Experimental Data Test Results}

Moving beyond perfect model tests, we next evaluate the method's capability on a more realistic condition that combines limited experimental bicrystal data with synthetic augmentation. For experimental simulations, we first approximate the prior distribution using a GMM and select a stress threshold for each grain boundary orientation.

\begin{figure}
    \centering
    \includegraphics[width=1\linewidth]{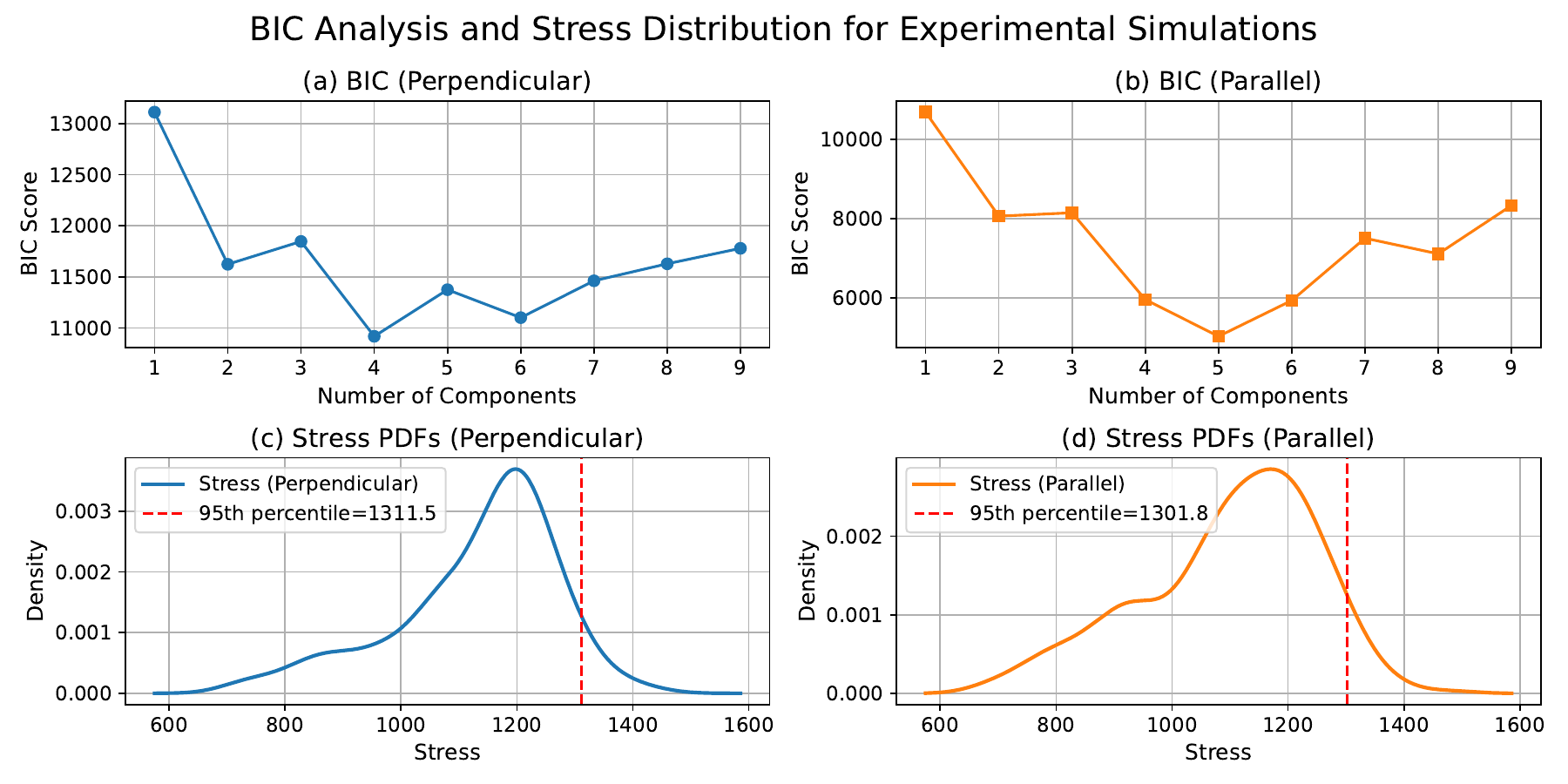}
    \caption{BIC analysis for GMM fitting of prior distribution $\mathbf{z}$ in the latent space, and stress distributions for for both bicrystal
configurations. Panel (a) and (b): Bayesian Information Criterion (BIC) as a function
of mixture components. Panel (c) and (d): Stress PDFs showing 95th percentile thresholds (red dashed lines), used to define extreme events.}
    \label{fig:bic_analysis}
\end{figure}

Figure \ref{fig:bic_analysis} shows the Gaussian component number selection for microstructural features and stress PDFs for both perpendicular and parallel bicrystal configurations. The BIC selects $K=4$ components for the perpendicular case and $K=5$ for the parallel case in Panels (\ref{fig:bic_analysis}a) and (\ref{fig:bic_analysis}b). The stress distributions in Panels (\ref{fig:bic_analysis}c) and (\ref{fig:bic_analysis}d) show similar non-Gaussian behaviors, with 95th-percentile thresholds of $\bar{\sigma} = 1311.5$ for the perpendicular configuration and $\bar{\sigma} = 1301.8$ for the parallel configuration. These thresholds are used to define the extreme events in the subsequent analysis. The selection of the 95th percentile as the threshold for extreme events stems from the physical hypothesis that the nucleation of damage in structural materials is an extreme-event process in a spatial sense \cite{dunham2025attribution, schmelzer2025statistical, bronkhorst2021local, lieberman2016microstructural}.

In our training process, we work with 300 experimental bicrystal samples alongside 700 synthetic samples from the fitted prior distribution. After 200 iterations, we test the model on the leftover experimental data, which is supplemented with additional synthetic samples. This assessment shows the practical utility of this method under data-scarce conditions typical of materials science applications.

Importantly, while inference is carried out in a reduced latent space, our evaluation emphasizes the physical feature space. For each sample, we have both the microstructural features and the corresponding stress state in original physical space, while the posterior distribution in the latent space provides complementary probabilistic information. By collecting the points with high posterior probability, we obtain a statistical analysis for the microstructural configurations that trigger extreme events.

\begin{figure}[htbp]
    \centering
    \includegraphics[width=1\linewidth]{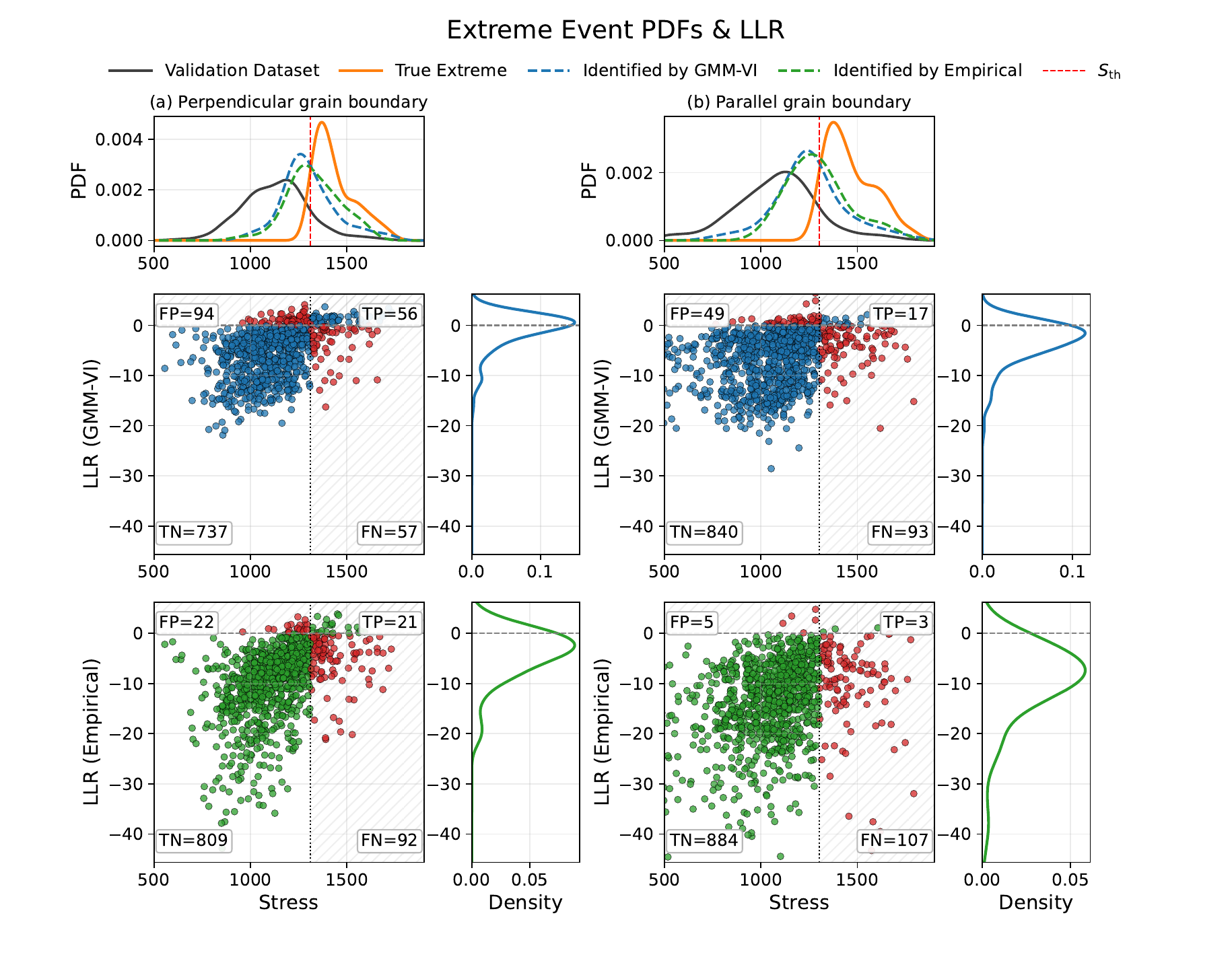}
    \caption{Performance evaluation using combined experimental bicrystal data and synthetic samples for both perpendicular (a) and parallel (b) grain boundary configurations. Top row shows probability density functions (PDFs) comparing the validation dataset (gray), true extreme events (orange), events identified by GMM-VI (blue dashed), events identified by empirical method (green dashed), and the stress threshold $S_{\text{th}}$ (black dashed vertical line). Middle and bottom rows display LLR scores versus stress values for GMM-VI and empirical methods respectively, with points colored by prediction correctness (red color represents incorrectness while blue and green color represent correctness). Side panels show the distribution of LLR scores. Classification metrics are provided for each method: TP (true positives), FP (false positives), TN (true negatives), and FN (false negatives).}
    \label{fig:stress_LLR_PDFs}
\end{figure}

Figure \ref{fig:stress_LLR_PDFs} presents performance evaluation for both perpendicular and parallel grain boundary configurations among test datasets.
The results in the top row reveal that both GMM-VI and empirical methods exhibit similar behavior. The stress PDFs of the identified extreme events, derived from both methods, capture the key shift away from the prior distribution and toward the actual extreme-event distribution (orange curve). However, the presence of inevitable misclassifications in both methods results in a distribution that does not perfectly align with the true exceedance. Notably, the empirical distribution shows fewer points exceeding the LLR threshold, resulting in unreliable classification.
The LLR distributions in the right panels provide additional perspective by showing the distributions of LLR values computed for actual extreme-event points. Here, the results from GMM-VI exhibit a more concentrated distribution of positive (or near-zero) LLR values, indicating greater ability to identify extreme events. In contrast, the empirical distribution exhibits a broader spread with many values below the LLR threshold. These differences demonstrate GMM-VI's ability to distinguish extreme events from normal events, as evidenced by its more decisive positive LLR assignments for actual extreme cases.
The LLR scatter plots in the middle and bottom rows confirm these trends, showing that GMM-VI maintains better performance across both grain boundary configurations. The consistency between these realistic mixed-data results and the perfect model tests validates the framework's robustness.

\begin{figure}[ht]
    \centering
    \includegraphics[width=1\linewidth]{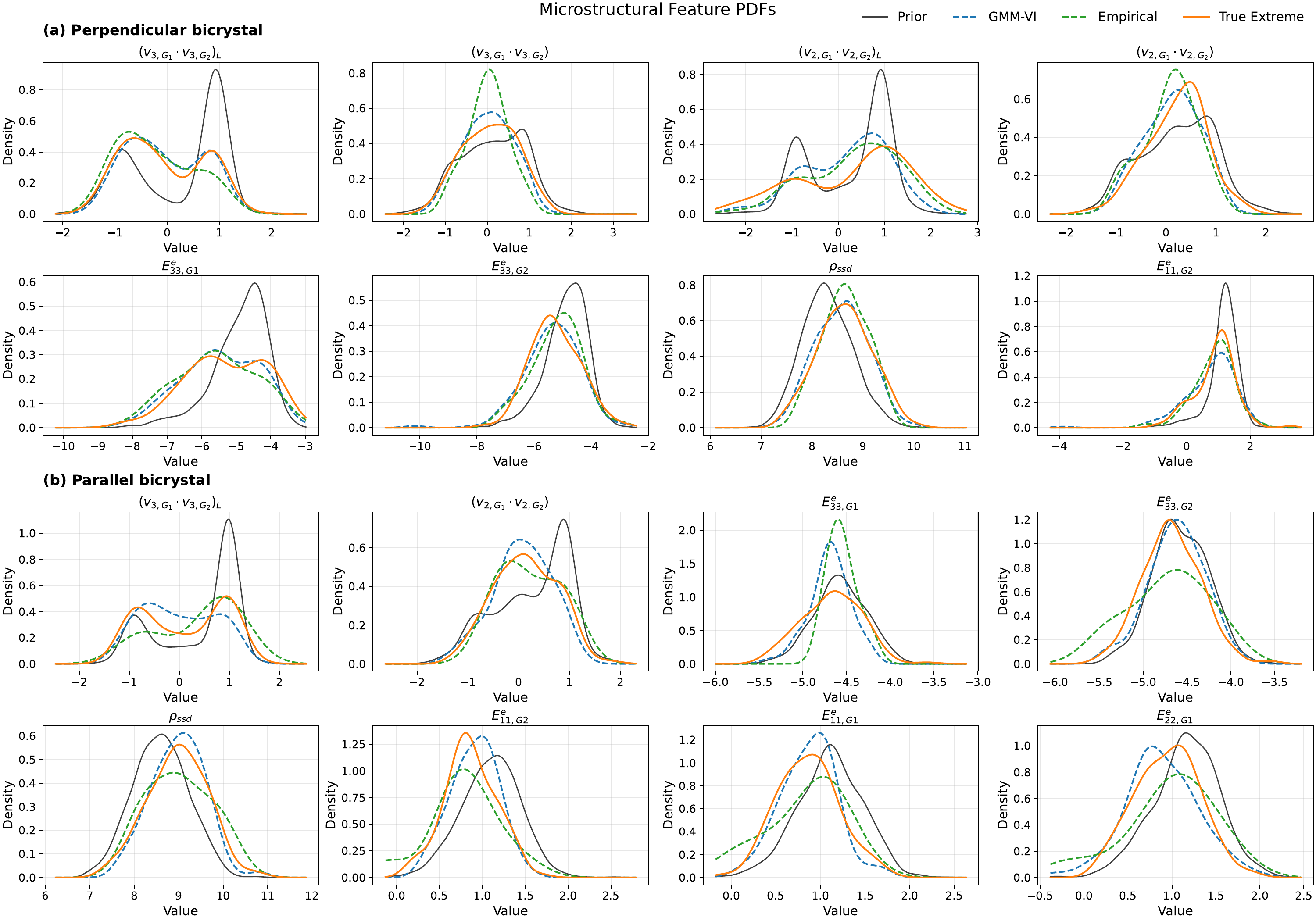}
    \caption{Probability density functions of selected microstructural features comparing prior distribution (gray), GMM-VI posterior (blue dashed), empirical distribution (green dashed), and true extreme events (orange) for (a) perpendicular and (b) parallel bicrystal configurations. Eight representative features are shown, including crystallographic descriptors, elastic strain components, elastic stiffness components, plastic strain eigenvalue, and other microstructural parameters. }
    \label{fig:Microstructural Feature PDFs}
\end{figure}

To more deeply understand extreme-event attribution from a physical perspective, we examine how inference methods capture distributional shifts in individual microstructural features.
Figure \ref{fig:Microstructural Feature PDFs} shows the part of microstructural feature distributions across both bicrystal configurations, which are selected based on PCA variance contributions discussed in Section \ref{Sec:PCA}. The comparison shows that GMM-VI posterior distributions (blue dashed) successfully concentrate around the true extreme-event distributions (orange) for key microstructural descriptors. For instance, certain crystallographic descriptors and elastic strain components exhibit clear shifts from the broad prior (gray) to focused posterior distributions that closely match the true extreme-event distributions.
The differences between perpendicular and parallel configurations are also evident at the feature level. In the perpendicular case, the distribution shifts are more pronounced, especially for the elastic strain components. This observation is consistent with the better classification performance shown in Figure \ref{fig:stress_LLR_PDFs}. These microstructural feature-level aspects provide physical interpretability to the statistical inference results, connecting the mathematical framework to underlying deformation mechanisms.

For example, let us analyze $E_{11,G2}^e$ (lower right part of both panels (\ref{fig:Microstructural Feature PDFs}a) and (\ref{fig:Microstructural Feature PDFs}b)). The extreme-event distribution predicted by the GMM-VI method (dashed blue curve) indicates that larger deformations that are parallel to the boundary result in stress localizations. Also note that $E_{33,G1}^e$ and $E_{33,G2}^e$ have qualitatively similar distributions, with a slight leftward shift, indicating an increase in elastic strain in the global compression direction. However, small differences in these distributions reveal that mismatched elastic strains across the boundary lead to extreme-stress events. Another interesting mechanism for stress localization is revealed by $(v_{3,G1} \cdot v_{3,G2})_L$. In the original numbering system for the principal components of the plastic stretching tensor as discussed in \cite{dunham2025attribution},  $v_{3,GN}$ is the principal compression direction. This quantity is closely aligned with the global compression direction, but not exactly so due to plastic anisotropy, i.e., the activation of specific slip systems by elevated resolved shear stresses. This implies that extreme stress events are usually triggered by a strong misorientation of the overall compression directions in the material, resulting in excess deformation at the grain interface. This is also exacerbated by elevated plastic flow, indicated by the increase in statistically stored dislocation density, $\rho_{ssd}$. These results strengthen the hypothesis that differences in the magnitudes of both grains' propensities to accommodate elastic and plastic deformations, as well as mismatches in their principal directions of deformation (both elastic and plastic), result in extreme stress events.

\section{Discussion}\label{Sec:Discussion}

In this study, we develop a new method to identify which microstructural features lead to extreme stress events in metals. Our method integrates Bayesian inference and a physics-based model. The former handles uncertainty in a principled way while the latter keeps our results grounded in realistic material behavior.
Several key components are incorporated to overcome the challenges of extreme event analysis. First, the PCA method makes inference tractable by projecting high-dimensional microstructural features into a low-dimensional latent space, while preserving essential variability and allowing reconstruction back to physical space. Second, GMMs then capture the non-Gaussian behavior in microstructural features reflecting the complexity of different mechanisms. Besides, the variational inference approach with responsibility weighting emphasizes the likelihood of extreme events during posterior updates. Additionally, the EVT further highlights the tail statistics. Furthermore, the physics-informed stress model provides mechanistic grounding by linking latent variables to measurable stress responses through established relationships in crystal plasticity.

We validate the framework through a perfect model test, where the physics-informed statistical model serves as a ground truth. The results demonstrate that the variational inference framework outperforms both the MCMC method and empirical approaches, achieving higher sensitivity in detecting extreme events while remaining computationally efficient. Mixed-model experimental tests are also conducted, combining limited bicrystal simulations with synthetic samples. The framework not only shows reliable classification performance but also offers physical insight. For example, it reveals that extreme stress events are primarily driven by mismatches in elastic strain across grain boundaries and by misalignment in the principal directions of plastic deformation between neighboring grains. These physically interpretable insights, quantitatively derived from the statistical posterior, bridge the gap between data-driven discovery and mechanistic understanding.

Several future directions could be pursued to extend this framework. On the methodological side, using nonlinear dimensionality reduction techniques like variational autoencoders would allow for richer latent representations to handle other complex systems while maintaining reconstruction ability. On the materials side, the current bicrystal validation provides a foundation. Future studies can be extended to more complicated systems, such as quad-crystal, octu-crystal, and larger polycrystalline configurations with many interacting grains. Systematic application across different grain types could provide a clear mapping of how failure mechanisms evolve with local grain structure.

\section*{Acknowledgment}
C.A.B. and N.C. are grateful for the support from NSF DMREF-CMMI 2118399. S.D.D. and Y.Z. are supported as research assistants under this grant.

\section{Appendix}

\subsection{Relationship between microstructural features and principal components}\label{Sec:PCA}

PCA method transforms the original microstructural features $\mathbf{x} \in \mathbb{R}^D$ into a lower-dimensional latent space $\mathbf{z} \in \mathbb{R}^d$ through the linear transformation: $$\mathbf{z} = \mathbf{V}^T(\mathbf{x} - \boldsymbol{\mu})$$,
where $\boldsymbol{\mu}$ is the feature mean vector and $\mathbf{V} = [\mathbf{v}_1, \mathbf{v}_2, \ldots, \mathbf{v}_d]$ contains the first $d$ eigenvectors of the feature covariance matrix, ordered by decreasing eigenvalues $\lambda_1 \geq \lambda_2 \geq \ldots \geq \lambda_d$. Each eigenvector $\mathbf{v}_k = [v_{1,k}, v_{2,k}, \ldots, v_{D,k}]^T$ defines the $k$-th principal component direction in the original feature space.

\begin{figure}[htbp]
    \centering
    \begin{subfigure}[b]{1\textwidth}
        \includegraphics[width=\textwidth]{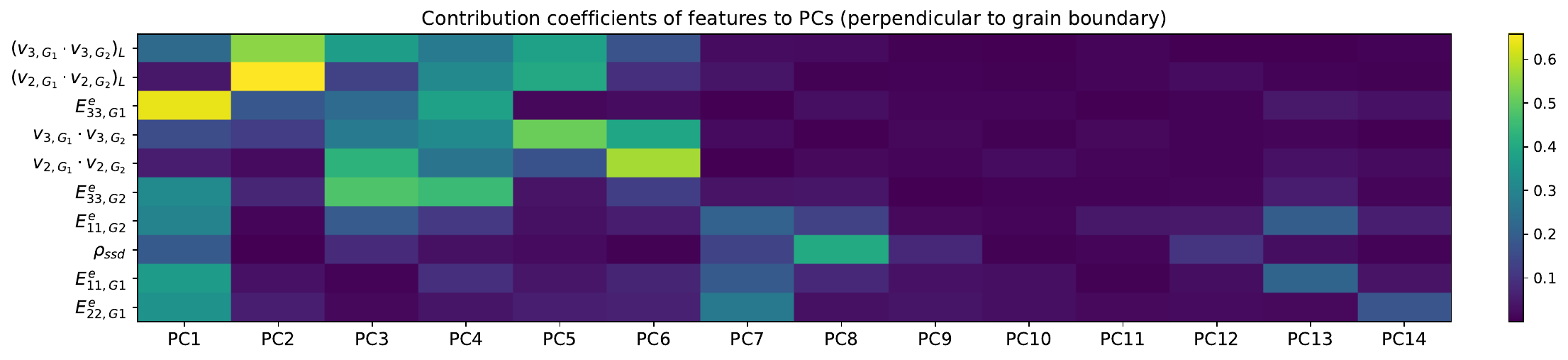}
        \caption{Bicrystal configuration simulations under perpendicular grain boundary.}
        \label{fig:a}
    \end{subfigure}
    \hfill
    \begin{subfigure}[b]{1\textwidth}
        \includegraphics[width=\textwidth]{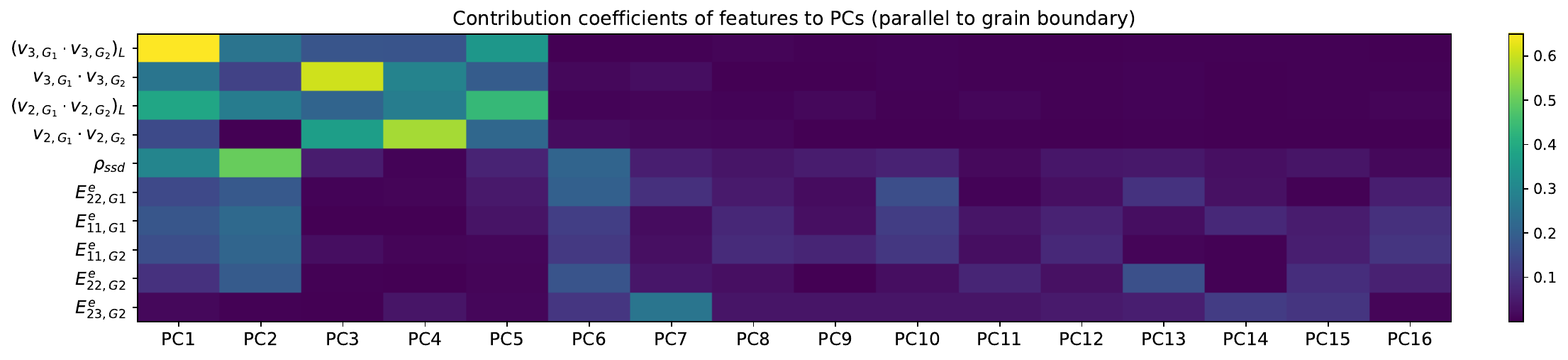}
        \caption{Bicrystal configuration simulations under parallel grain boundary.}
        \label{fig:b}
    \end{subfigure}
    \caption{Contribution coefficients of microstructural features to principal components (PCs). Each heatmap shows the absolute values of coefficients $|v_{j,k}\sqrt{\lambda_k}|$, where $v_{j,k}$ is the $j$-th component of the eigenvector for PC$_k$ and $\lambda_k$ is the corresponding eigenvalue. Brighter colors indicate stronger contributions of feature $j$ to PC$_k$.}
    \label{fig:pca_analysis}
\end{figure}

To provide interpretability for the PCA-transformed latent space used in our variational inference framework, we analyze the contribution of individual microstructural features to the principal components. Figure \ref{fig:pca_analysis} presents contribution coefficient heatmaps for both bicrystal configurations. These heatmaps display the top 10 microstructural features based on their contributions to all principal components. Here, the contributions are measured by the absolute value of the weighted eigenvector coefficient $|v_{j,k}\sqrt{\lambda_k}|$. These PCA contribution maps highlight which physical features dominate the all principle components. In the perpendicular configuration, elastic strain components appear most prominently, whereas in the parallel configuration, stiffness components and dislocation density contribute more strongly. PCA contributions provide a direct mapping between latent principal components and interpretable physical features. Importantly, because PCA is a linear and reversible transformation, the identified combinations of features in latent space can be reconstructed back into the original microstructural descriptors. This ensures that any posterior shifts observed in the latent space translate into trackable changes in elastic stiffness, strain, dislocation density, or misorientation features, and further provides information with us.

\subsection{Derivation of Responsibility Updates}\label{appendix:Derivation}

The responsibilities $r_{ik}$ are derived by maximizing the ELBO with respect
to the assignment probabilities. Recall that the ELBO is defined as:
\begin{equation}
\text{ELBO} = \mathbb{E}_Q[\log P(\mathbf{E} \mid \mathbf{z})]
- \text{KL}(Q(\mathbf{z}) \| P(\mathbf{z})).
\end{equation}

Expanding the ELBO in terms of the mixture components and responsibilities:
\begin{equation}
\begin{aligned}
\text{ELBO} &= \sum_{i=1}^{N_E} \sum_{k=1}^K r_{ik} \log P(\sigma_i > \bar{\sigma} \mid \mathbf{z}_i) \\
&\quad + \sum_{i=1}^{N_E} \sum_{k=1}^K r_{ik} \left[\log \omega_k
- \frac{1}{2}\log|\Lambda_k| - \frac{1}{2}(\mathbf{z}_i - \nu_k)^\top \Lambda_k^{-1}(\mathbf{z}_i - \nu_k)\right] \\
&\quad - \sum_{i=1}^{N_E} \sum_{k=1}^K r_{ik} \left[\log \pi_k
- \frac{1}{2}\log|\Sigma_k| - \frac{1}{2}(\mathbf{z}_i - \mu_k)^\top \Sigma_k^{-1}(\mathbf{z}_i - \mu_k)\right] \\
&\quad - \sum_{i=1}^{N_E} \sum_{k=1}^K r_{ik} \log r_{ik},
\end{aligned}
\end{equation}
where the last term is the entropy of the categorical distribution over component
assignments.

To maximize the ELBO with respect to $r_{ik}$ subject to the normalization
constraint $\sum_{k=1}^K r_{ik} = 1$ for each sample $i$, we set the derivative
to zero. This yields the unnormalized responsibility:
\begin{equation}
\begin{aligned}
\log \tilde{r}_{i, k} & =\log \omega_k-\frac{1}{2} \log \left|\boldsymbol{\Lambda}_k\right|-\frac{1}{2}\left(\mathbf{z}_i-\boldsymbol{\nu}_k\right)^\top \boldsymbol{\Lambda}_k^{-1}\left(\mathbf{z}_i-\boldsymbol{\nu}_k\right) \\
& +\log P\left(\sigma_i > \bar{\sigma} \mid \mathbf{z}_i\right)-\left[\log \pi_k-\frac{1}{2} \log \left|\boldsymbol{\Sigma}_k\right|-\frac{1}{2}\left(\mathbf{z}_i-\boldsymbol{\mu}_k\right)^\top \boldsymbol{\Sigma}_k^{-1}\left(\mathbf{z}_i-\boldsymbol{\mu}_k\right)\right],
\end{aligned}
\end{equation}
which can be written compactly as:
\begin{equation}
\tilde{r}_{ik} \propto \frac{\omega_k \mathcal{N}(\mathbf{z}_i \mid \nu_k, \Lambda_k)
\cdot P(\sigma_i > \bar{\sigma} \mid \mathbf{z}_i)}
{\pi_k \mathcal{N}(\mathbf{z}_i \mid \mu_k, \Sigma_k)}.
\end{equation}

The responsibilities are then normalized:
\begin{equation}
    r_{i,k} = \frac{\tilde{r}_{i,k}}{\sum_{j=1}^K \tilde{r}_{i,j}}.
\end{equation}

\bibliography{references}

\end{document}